\documentclass[letterpaper,traditabstract]{aa} 
\usepackage{graphicx}
\usepackage{natbib}
\usepackage[hyperindex,breaklinks]{hyperref}
\bibpunct{(}{)}{;}{a}{}{,} 
\begin{document}
%
\title{Transit spectrophotometry of the exoplanet HD189733b.  I.  Searching for water but finding haze with HST NICMOS}

\author{David K. Sing\inst{1}, J.-M. D\'esert\inst{1}, A. Lecavelier des Etangs\inst{1}, G. E. Ballester\inst{2}, A. Vidal-Madjar\inst{1}, V. Parmentier\inst{1}, G. Hebrard\inst{1}, \& G. W. Henry\inst{3}}

   \offprints{D. K. Sing}

\institute{UPMC Univ Paris 06, CNRS, Institut d'Astrophysique de Paris, 98bis boulevard Arago, F-75014 Paris, France\\ \email{sing@iap.fr}
 \and
           Lunar and Planetary Laboratory, University of Arizona, 
           Sonett Space Science Building, Tucson, AZ 85721-0063, USA
 \and
          Center of Excellence in Information Systems, Tennessee State University, 
          3500 John A. Merritt Blvd., Box 9501, Nashville, TN 37209, USA
}

   \date{Received 29 June 2009 / Accepted 27 July 2009}

  \abstract{
We present Hubble Space Telescope near-infrared transit photometry of the nearby hot-Jupiter \object{HD189733b}.  The observations were taken with the NICMOS instrument during five transits, with three transits executed with a narrowband filter at 1.87 $\mu$m and two performed with a narrowband filter at 1.66 $\mu$m.  Our observing strategy using narrowband filters is insensitive to the usual HST intra-orbit and orbit-to-orbit measurement of systematic errors, allowing us to accurately and robustly measure the near-IR wavelength dependance of the planetary radius.  Our measurements fail to reproduce the Swain et al. absorption signature of atmospheric H$_2$O below 2$\mu m$ at a 5$\sigma$ confidence level.  We measure a planet-to-star radius contrast of 0.15498$\pm$0.00035 at 1.66 $\mu$m and a contrast of 0.15517$\pm$0.00019 at 1.87 $\mu$m.  Both of our near-IR planetary radii values are in excellent agreement with the levels expected from Rayleigh scattering by sub-micron haze particles, observed at optical wavelengths, indicating that upper-atmospheric haze still dominates the near-IR transmission spectra over the absorption from gaseous molecular species at least below 2 $\mu m$.
}

\titlerunning{HST/NICMOS Transit Spectrophotometry of HD189733b}
\authorrunning{Sing et. al}
   \maketitle

\section{Introduction}

Transiting close-in exoplanets have provided the extraordinary possibility to begin studying the detailed characteristics of extrasolar planets.  A transit or anti-transit event allows for an exoplanet to be temporally resolved from the bright parent host star, isolating such valuable information as the planet's absorption or emission spectra.  During a primary transit event, both the opaque body of the planet as well as its atmosphere blocks light from the parent star.  A precise determination of the radius of the planet can be made from the total obscuration, while partial transmission of light through the exoplanet's atmosphere, with its wavelength and altitude dependence, allows for detecting composition and structure. The frequent transits, anti-transits, and large signals make hot Jupiters the best targets for these studies. Space-based observatories, notably HST and Spitzer, have proven to be extremely efficient platforms on which to do follow-up transit and anti-transit studies given their superior photometric performance, wide wavelength range, and ability to observe multiple transits over long baselines within a single observing season.

Two hot Jupiters in particular, \object{HD189733b} and \object{HD209458b}, currently offer the very best laboratories in which to study exoplanet atmospheres and have become the prototype hot-Jupiter planets. These two planets have the brightest parent stars among transiting planets and have large transit depths, making precise studies at high signal-to-noise ratios possible.  The first transiting planet discovered, HD209458b, holds the distinction of the first detection of an extrasolar planetary atmosphere \citep{2002ApJ...568..377C} and escaping atmosphere \citep{2003Natur.422..143V,2004ApJ...604L..69V, 2008ApJ...676L..57V, 2007Natur.445..511B}.  The optical transmission spectra of this planet shows evidence for several different layers of Na \citep{2008ApJ...686..658S,2008ApJ...686..667S} as well as Rayleigh scattering by molecular hydrogen \citep{2008A&A...485..865L} and the presence of TiO/VO \citep{2008A&A...492..585D}.  The atmospheric Na signature has also been confirmed by ground based observations \citep{2008A&A...487..357S}. This planet also exhibits a stratospheric temperature inversion \citep{2007ApJ...668L.171B,2008ApJ...673..526K} thought to be caused by a strong optical absorber such as TiO \citep{2008A&A...492..585D,2003ApJ...594.1011H,2008ApJ...678.1419F} and/or HS and S$_2$ \citep{2009ApJ...701L..20Z}.

HD189733b \citep{2005A&A...444L..15B, 2006A&A...445..341H} is among the closest known transiting planets with a K1.5V type parent star, giving it one of the largest transit and anti-transit signals known.  Spitzer anti-transit measurements have shown efficient heat redistribution, measuring the planet's temperature profile from orbital phase curves \citep{2007Natur.447..183K,2009ApJ...690..822K} and a definitive detection of atmospheric water from emission spectra \citep{2008Natur.456..767G}.  The signature of CO$_2$ has also been seen from HST/NICMOS emission spectra \citep{2009ApJ...690L.114S}.  In primary transit, \cite{2008MNRAS.385..109P} used the ACS grism to provide the first transmission spectra from 0.6 to 1.0 $\mu$m. This spectrum is seen to be almost featureless likely indicating the presence of high altitude haze, with a $\lambda^{-4}$ wavelength dependence of the spectra likely due to Rayleigh scattering by sub-micron MgSiO$_3$ molecules \citep{2008A&A...481L..83L}.  From the ground, \cite{2008ApJ...673L..87R} detected strong Na absorption in the core of the doublet.  In the infrared, \cite{2008Natur.452..329S} used HST/NICMOS and showed evidence for absorption of atmospheric water and methane while Spitzer broadband transit photometry has given evidence for CO \citep{2009ApJ...699..478D}.
  
The current transmission spectrum of HD189733b appears to be one which is dominated by condensed haze particles in the optical, but dominated by the absorption of gaseous molecular species (e.g., H$_2$O, CO, CH$_4$) at near-infrared and infrared wavelengths.  The transition wavelength between being dominated by haze or gaseous species should give indications of the particle size of the condensed haze, though it is not clear if all the current observations are completely compatible and can give a coherent picture of the planet's atmosphere.  

Here we present near-infrared HST photometry of HD189733b during five transits, designed with the intended goal of measuring the signature of atmospheric water.  
This work is part of our ongoing efforts to charaterize the transit spectra of HD189733 using space-based observatories \citep{2009ApJ...699..478D,2009submittedD}.  We observed with two narrowband filters and designed the observing program to be unbiased from the typical orbit-to-orbit and intra-orbit systematic errors inherent in observing in spectroscopic modes with HST STIS, ACS (e.g., \citealt{2001ApJ...552..699B,2007A&A...476.1347P}) and NICMOS (e.g., \citealt{2008Natur.452..329S,2009ApJ...696..241C}) instruments.  In Sect. 2 and Sect. 3, we detail these observations, which include correcting for the presence of unocculted star spots and stellar limb darkening.  In Sect. 4 we present the analysis of our transit light curves and discuss our results in Sect. 5 and Sect. 6.

\begin{figure}
 {\centering
  \includegraphics[width=0.36\textwidth,angle=90]{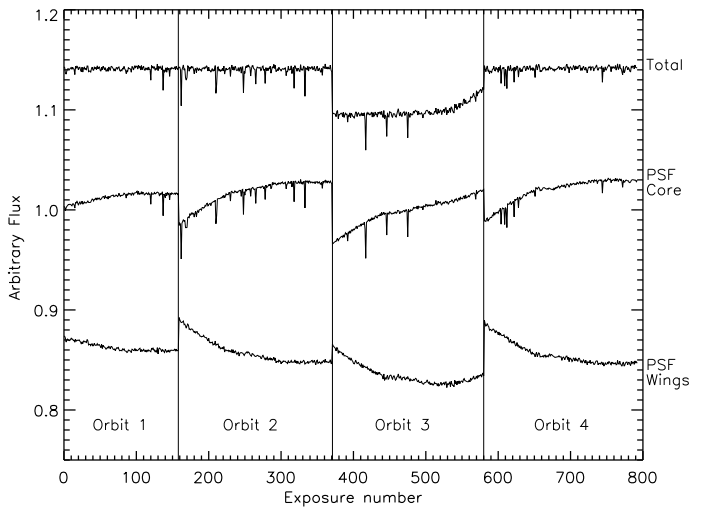}}
\caption[]{\footnotesize{Aperture photometry from Visit \#1 using the F187N narrowband
 filter.  The data from all four spacecraft orbits of the HST visit are
 plotted back-to-back (there are $\sim$45 minute data gaps between HST
 orbits due to Earth occultation).  The transit occurs in orbit 3
 between exposures 371 and 580 and is visible in the top plot.  Plotted
 are the total counts from a 15.2 pixel aperture (top), along with the
 counts from only the PSF central core (middle, 4.25 pixel aperture),
 and counts from only the PSF wings containing the
1$^{st}$ and 2$^{nd}$ airy rings (bottom, ring with inner-annulus of 4.25 pixels outer-annulus 15.2 pixels).  The intra-orbit thermal focus variations change the shape of the PSF during an orbit, which can be seen by the large flux changes in counts of the PSF core and wings.  However, with our narrowband filters the total counts are conserved in wide apertures, which exhibit no residual intra-orbit variations.  A small intrapixel sensitivity related error is observed in the PSF core photometry which contaminates $\sim$10\% of the exposures (see text).}}
\end{figure}

\section{Observations}
\subsection{Hubble Space Telescope NICMOS narrowband photometry}
We observed HD189733 during five transits using the Near Infrared Camera and Multi Object Spectrometer (NICMOS) aboard the Hubble Space Telescope during Cycle 16 (GO-11117).  Each transit observation consisted of four consecutive spacecraft orbits, each roughly centered on a transit event.  The visits occurred during: 15 April 2008 UT 15:36 to 21:17 for visit \#1; 3 May 2008 UT 08:51 to 14:26 for visit \#2; 18 May 2008 UT 22:44 to 19 May 2008 UT 04:27 for visit \#3; 25 May 2008 UT 14:33 to 20:15 for visit \#4; and 11 Aug 2008 UT 06:24 to 12:04 for visit \#5.  For each transit event, we obtained images using the high resolution NIC1 camera with only a single narrowband filter, using either the F166N filter centered at 1.6607 $\mu$m ($\Delta\lambda=0.0170\mu$m) or the F187N filter centered at 1.8748 $\mu$m ($\Delta\lambda=0.0191\mu$m).  By not switching between filters, the duty cycle of the instrument is increased as much less overhead time is spent between consecutive exposures.

The NICMOS instrument has three 256x256 pixel HgCdTe actively cooled cameras with the NIC1 camera offering the highest spatial resolution, having a 11$\times$11 arcsec field of view and 43 milliarsec sized pixels.  The high resolution of the camera combined with the use of narrowband filters allowed us to observe the bright (V$_{mag}$=7.67) star with short 2-4 second exposures well before saturating the detector.  We also adopted a modest defocus during our observations, such that the photometry would be more insensitive to flat fielding errors and sub-pixel variations.
The F187N filter images were taken in Multiaccum SPARS4 mode with Nsamp=3 yielding 4.00 second exposures while the F166N filter images were taken with Multiaccum STEP1 mode with Nsamp=4 yielding 1.99 second exposures.  Both observing modes have a 9 second readout overhead.  In visit \#1 we obtained 793 F187N exposures, in visit \#2 860 F166N exposures, in visit \#3 418 F187N exposures, in visit \#4 927 F166N exposures, and in visit \#5 751 F187N exposures.  The data quality of visit \#3 was compromised by a failure of a known 2-gyro issue with the Fixed Head Star Trackers.  The failure caused a total loss of the visit \#3 out-of-transit data.  However, even without a long out-of-transit baseline flux, visit \#3 still proved useful when analyzing the data in conjunction with the other two F187N visits.

During each HST visit, we obtained 50 exposures in one filter during the beginning of the first orbit, before switching to the opposite narrowband filter in which the rest of the visit, including the transit, was observed.  This short set of opposite-band images was used to help monitor the stellar activity, such that the absolute flux and activity level of the star would be known at each of our two wavelengths for all five visits.
   
We performed aperture photometry on the calibrated STScI pipeline reduced images.  The pipeline includes corrections for bias subtraction, dark current, detector non-linearities, and applies a flat field calibration.  The aperture location for each image was determined using a two dimensional Gaussian fit of the point spread function (PSF) for each image.  We found apertures with a radii of 15.2 pixels for the F187N filter and 14.7 pixels for the F166N filter minimized the standard deviation of the out-of-transit light curves.  The background in our short exposures was found to be negligible, typically accounting for only $\sim$50 total counts per image.

Our choice of a wide aperture minimizes the effect of intra-orbit
thermal focus variations (see Fig. 1).  This well known effect (see STScI Instrument Science Reports
ACS2008-03\footnote{http://www.stsci.edu/hst/acs/documents/isrs/} and NICMOS ISR2007-003\footnote{http://www.stsci.edu/hst/nicmos/documents/isrs/}) changes the shape of the PSF during an orbit and is due to
small changes in the focus of the telescope, which are ascribed to
temperature variations.  With a sufficiently wide aperture, covering at
least the central peak and first and second airy rings, no residual
intra-orbit variations are observed.  In four of our five orbits, the photometric time series exhibit sudden drops in flux,
confined to the central core of the PSF and effects 10-15\% of the
images.  We attribute this phenomenon to a likely intrapixel sensitivity
variation, as it is confined to pixels in the central PSF core which
have the largest change in flux across a pixel and the effect is smaller
for higher levels of defocus.  Images which show this intrapixel
sensitivity effect were easily flagged by comparing the photometric time
series of the photometry of the central PSF core to the photometry of
the outer-airy rings, which are insensitive to the effect (see Fig. 1).  We adopted two strategies to deal with these flagged images, choosing to either eliminate them from the remaining analysis or interpolate the affected PSF core photometry using neighboring images in the time series.  Both methods produced equivalent results in the final transit light curve fits.  We note that future HST/NICMOS programs can eliminate this small problem completely by adopting a larger defocus, as is the case here in visit \#5 which does not exhibit this effect.

\subsection{Monitoring the stellar activity}
 We monitored the stellar activity of HD189733 using both ground-based data and the absolute flux level from the NICMOS instrument itself.  The ground-based coverage was provided by the T10 0.8 m Automated Photoelectric Telescope (APT) at Fairborn Observatory in southern Arizona.  This ongoing observing campaign of HD189733 began in October 2005 and is detailed in \cite{2008AJ....135...68H}.  The APT uses two photomultiplier tubes to simultaneously gather Stromgren $b$ and $y$ photometry.  The APT data spans four of our five HST visits (see Fig. 2), with only the fifth visit not covered as the telescope is shut down during the Arizona monsoon season.  This dataset also covers the HD189733 stellar activity during the epochs of the HST transmission spectra from \cite{2007A&A...476.1347P} and \cite{2008Natur.452..329S} as well as the Spitzer transit photometry of \cite{2009ApJ...699..478D}, \cite{2009submittedD}, and \cite{2009ApJ...690..822K} making it an invaluable resource when comparing the stellar activity level from epoch to epoch.

The near-IR flux levels for HD189733, during our five visits, are based on
the baseline flux level from each transit light curve as well as the
first 50 exposures obtained in the filter opposite to that of the transit
(ie. F166N or F187N).  While these first 50 exposures were obtained during the first
orbit of a visit, these orbits did not show any significant signs of systematic
errors.  Our use of the first orbit is in contrast to nearly all previous similar HST studies
(including the STIS and NICMOS instruments), which have had to disregarded the first orbit of an HST visit.  During the first orbit, the telescope thermally relaxes in its new pointing position causing significant flux variations.
The near-IR stellar flux levels are shown in Fig. 2, with an arbitrary zero-magnitude flux, which clearly
modulate in phase with the optical measurements.

\begin{figure}
 {\centering
  \includegraphics[width=0.36\textwidth,angle=90]{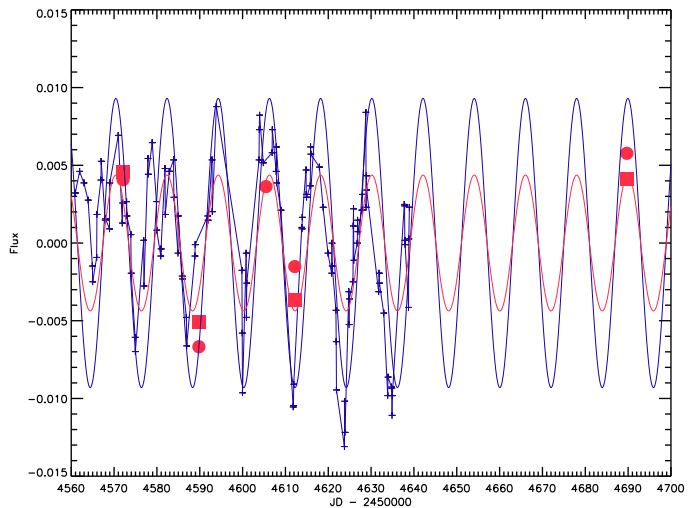}}
\caption[]{\footnotesize{Simultaneous ground-based photometric monitoring of HD189733 (blue crosses) continued from \cite{2008AJ....135...68H}, along with a fitted sinusoid (blue).  The NICMOS flux levels for all five visits are also plotted for both the F166N filter (red squares) and the F187N filter (red dots).  Assuming 1000\,K cooler star spots with $\sim2\%$ coverage, the amplitude of the near-IR spot modulation is expected to be about half that observed in the optical (red sinusoid).  The NICMOS flux levels are seen to modulate in phase with the optical ground-based data at approximately the expected near-IR spot amplitude.}}
\end{figure}

\section{Stellar limb darkening}
\subsection{Three parameter non-linear limb-darkening law}
For solar-type stars at near-infrared and infrared wavelengths, the
strength of stellar limb darkening is weaker compared to optical
wavelengths.  However, the intensity distribution is increasingly
non-linear at these longer wavelengths (see Fig. 5 of \citealt{2008ApJ...686..658S}), which can require adopting non-linear limb-darkening laws when fitting high S/N transit light curves.  Many studies have chosen to use a quadratic or four parameter non-linear limb-darkening law to describe the stellar intensity distribution \citep{2000A&A...363.1081C}, with the coefficients determined from stellar atmospheric models.  As noted by \cite{2008ApJ...686..658S} for the widely used Kurutz 1D ATLAS stellar models, the largest differences between existing limb-darkening data and the 1D stellar models is at the very limb, where ATLAS models predict a dramatic increase in the strength of limb darkening.  For the sun, the ATLAS models over-predict the strength of limb darkening by $>$20\% at $\mu=cos(\theta)$ values below 0.05, though they perform well otherwise.  Both solar data \citep{1994SoPh..153...91N} and 3D stellar models \citep{2006A&A...446..635B} show the intensity distribution at the limb to vary smoothly to $\mu=0$, with no dramatic or sudden increases in limb-darkening strength.  To mitigate this limb effect, we choose to fit for the limb-darkening coefficients from ATLAS models using only values of $\mu \geq 0.05$.  We also choose to fit a three parameter non-linear limb-darkening law,
\begin{equation}  \frac{I(\mu)}{I(1)}=1 - c_2(1 - \mu) - c_3(1 - \mu^{3/2}) - c_4(1 - \mu^{2}) \end{equation}
as the $\mu^{1/2}$ term from the four parameter non-linear law mainly affects the intensity distribution at small $\mu$ values and is not needed when the intensity at the limb is desired to vary linearly at small $\mu$ values.  Compared to the quadratic law, the added $\mu^{3/2}$ term provides the flexibility needed to more accurately reproduce the stellar model atmospheric intensity distribution at near-infrared and infrared wavelengths. 

\subsection{HD189733 limb darkening}
The limb-darkening coefficients for the three parameter non-linear law were computed using a Kurucz ATLAS stellar model\footnote{http://kurucz.harvard.edu/} with T$_{eff}$=5000 K, log g=4.5, and [Fe/H]=0.0 in conjunction with the transmission through our two narrowband filters.  For reasons described above, we fit for the limb-darkening coefficients using the calculated intensities between $\mu$=0.05 and $\mu$=1, providing the coefficients in Table~1.

\begin{table} 
\centering
\caption{Three parameter non-linear limb darkening coefficients}
\label{table:1}
\begin{tabular}{lll}
\hline\hline  
coefficient    &  F166N filter  &  F187N filter   \\ 
\hline 
 $c_2$         &  ~2.1483        &  ~1.9508        \\ 
 $c_3$         & -2.7763        & -2.4507          \\ 
 $c_4$         &  ~1.1265        &  ~0.9770        \\ 
\hline
\end{tabular}
\end{table}


\section{Reduction and analysis}
\subsection{Correcting for non-occulted stellar spots}
A transit light curve is affected by the presence of star spots, both
when a spot is occulted by the planet and by the presence of
non-occulted spots during the epoch of the transit observations
\citep{2008MNRAS.385..109P}.  We will focus on non-occulted spots here,
as there is no evidence for occulted spots in any of our observed
transit light curves.

Non-occulted stellar spots affect the shape of a transit light curve by
making the stellar surface dimmer than a spot-free stellar disk.  During
transit, the planet hides a larger fraction of the total overall flux,
thereby increasing the apparent size of the planet.  A wavelength
dependance is also introduced, as the stellar flux lost by the spots at
a given wavelength will depend on the blackbody temperature difference
between the stellar surface and the spots.  The stellar spots also
create a quasi-periodic photometric variability, as the spots rotate
into and out of view at the stellar rotation period.  

\begin{figure}
 {\centering
  \includegraphics[width=0.395\textwidth,angle=90]{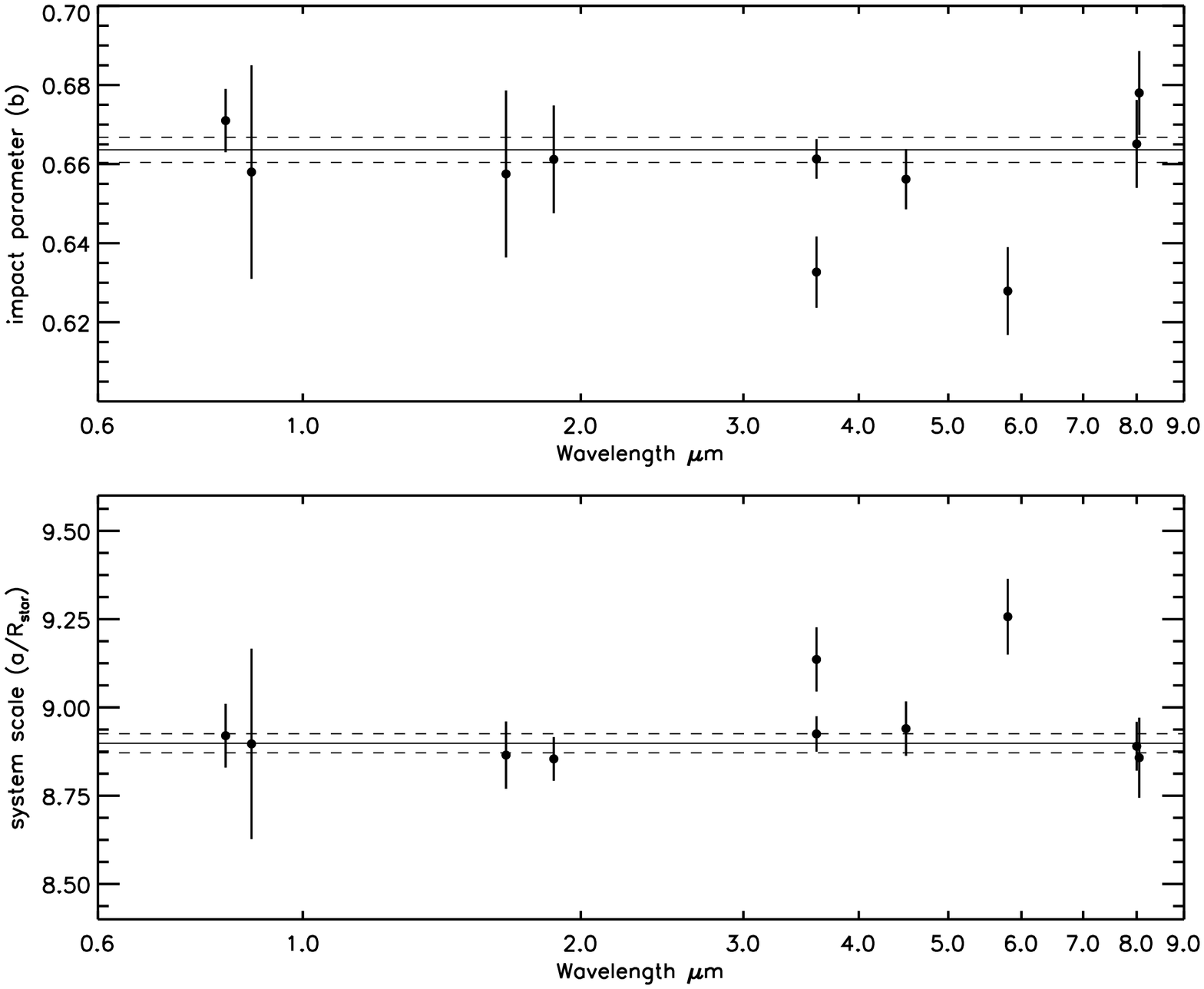}}
\caption[]{\footnotesize{The impact parameter (top) and system scale parameter (bottom) vs. wavelength for HD189733 as measured by \cite{2009ApJ...699..478D}, \cite{2009submittedD}, \cite{2009ApJ...690..822K}, \cite{2007A&A...476.1347P}, \cite{2007AJ....133.1828W}, and this work.  Excluding two deviant Spitzer values  from Desert et al. (2009; see also Paper II \citealt{2009submittedD} for further details), there is a good general agreement between the various HST, Spitzer, and ground-based results with a mean $b$ value of 0.6636$\pm$0.0031 and mean $a$/R$_{\star}$ value of 8.898$\pm$0.027.  These mean values are over-plotted (horizontal line) along with their 1$\sigma$ errors (dashed vertical lines) and adopted for our final transit light curve fits seen in Figures 5 and 6.}}
\end{figure}

To correct for the presence of non-occulted stellar spots in this
program, we used the ongoing ground-based photometric
monitoring data of \cite{2008AJ....135...68H} as well as the calibrated
flux level of the NICMOS observations, measuring the flux variation due
to stellar spot modulation.  HD189733 is an active star, varying in the
optical at the 2-3\% level with spot-coverage at any given epoch
covering $\sim$1-2\% of the stellar surface.  The NICMOS
observations vary by $\sim$1\% (see Fig. 2), which is consistent with
the expected near-IR variation due to stellar spots as it occurs both
in phase with the ground-based photometric monitoring and near the
amplitude expected when assuming $\sim$4000\,K spot temperatures as measured by
\cite{2008MNRAS.385..109P} and the observed optical spot coverage.
Furthermore, for each visit both the F166N and F187N filters showed
similar near-IR stellar flux variations, as would be expected for
stellar spots.  The baseline flux of each visit was corrected in a
differential manner, adding the necessary flux to each visit such that
all the observations are compared at the same minimum spot activity
baseline flux level.  HST visits \#1 and \#5 occurred at a spot activity
minimum, and were chosen here as the flux reference for the remaining
visits.  The corresponding effect on the final determined radius is
somewhat small but these corrections are important as exoplanetary atmospheric
signatures are typically also small.  An estimated $\sim$0.0015 uncertainty on the absolute flux level of each visit translates into a spot-correction uncertainty of $\sim$0.0001 R$_{pl}$/R$_{\star}$ and $\sim$0.25$\sigma$ on the final values of R$_{pl}$/R$_{\star}$ for the two wavelengths when combining the different visits.

\subsection{Transit light curve fits}
  We modeled the transit light curve with the theoretical transit models of \cite{2002ApJ...580L.171M}.  We choose to fix the planetary orbital phase using the ephemeris of \cite{2009ApJ...690..822K} (P = 2.21857578 $\pm$ 0.00000080 days; Tc = 2454399.23990 $\pm$ 0.00017 HJD), fitting for the planet-to-star radius contrast R$_{pl}$/R$_{\star}$, the inclination $i$, and the mean stellar density $\rho_{\star}$, which is proportional to the system scale ($a$/R$_{\star}$) cubed.  No significant deviations from the \cite{2009ApJ...690..822K} ephemeris were found, with the three F187N visits determined to be within 2 seconds of the predicted central transit time.  To account for the effects of limb-darkening on the transit light curve, we adopted the three parameter limb-darkening law and let the linear coefficient, $c_2$, free in the fit, while $c_3$ and $c_4$ were fixed to the best-fit model values listed in Table 1.  Choosing to fit only one limb-darkening parameter ensures that the fits are both not significantly biased by adopting a stellar atmospheric model and do not suffer from degeneracies between fitting multiple limb-darkening parameters.  We also allowed the flux level of each visit to vary in time linearly, described by two fit parameters.  The linear trend accounted for any possible detector drifts or stellar activity related flux variations during the observations and we found that visits \# 4 \& 5 exhibited a small slope.  The best fit parameters were determined simultaneously with a Levenberg-Marquardt least-squares algorithm using the unbinned data (see Table 2), with the error of each datapoint set to the standard deviation of the out-of-transit residuals of a given visit.  
\begin{figure}
 \centering
   \includegraphics[width=0.35\textwidth,angle=90]{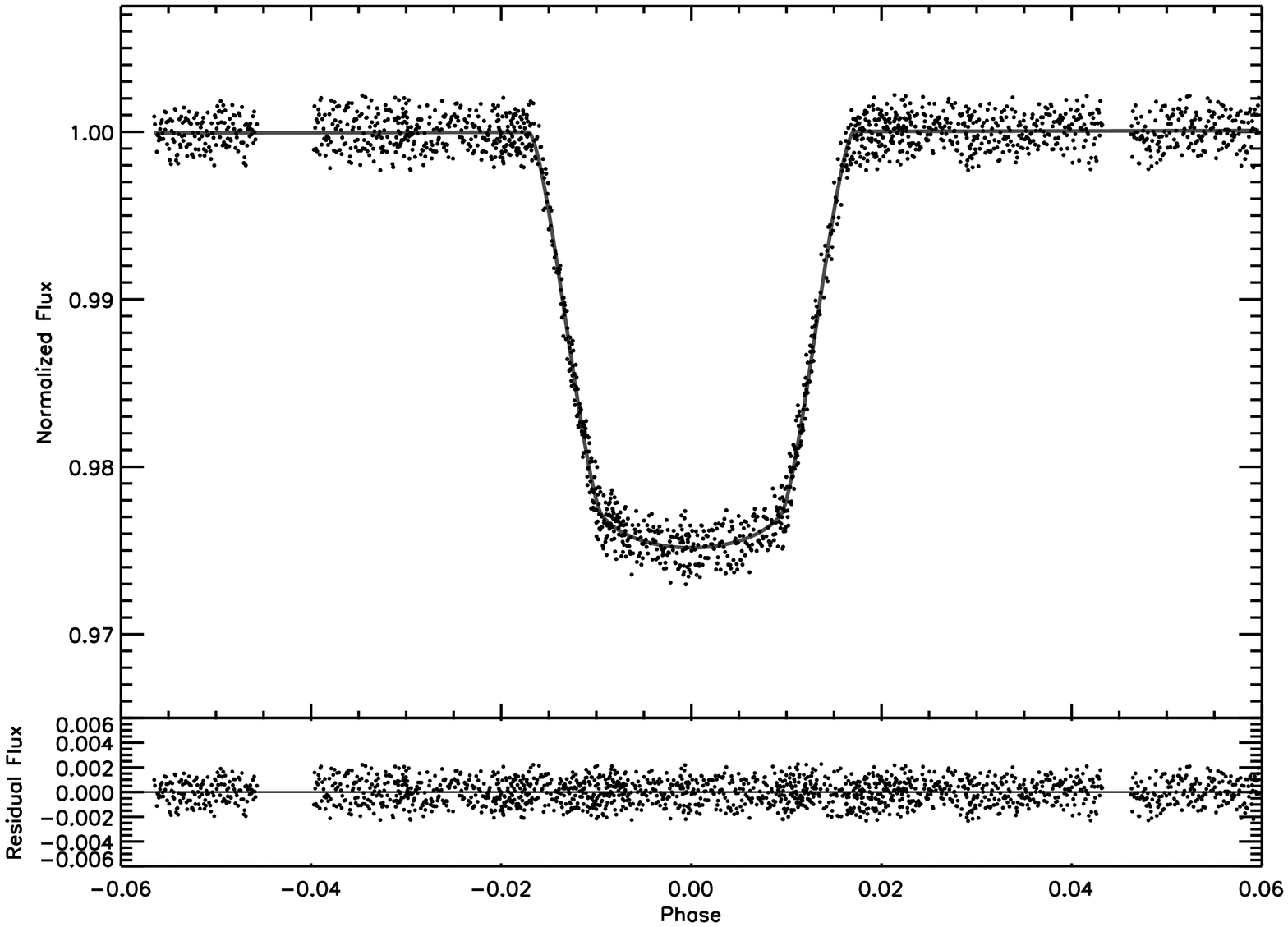}
\caption[]{\footnotesize{NICMOS filter transit photometry of all three visits using the 1.8748$\mu m$ narrowband filter.  The best-fit transit light curve model is also fit (grey) along with the observed - calculated residuals of the fit (bottom panel).  The light curve has a S/N of 1000 per-point and an equivalent precision of 4.0$\times10^{-4}$ in an 80 second bin.\\}}
 \centering
  \includegraphics[width=0.35\textwidth,angle=90]{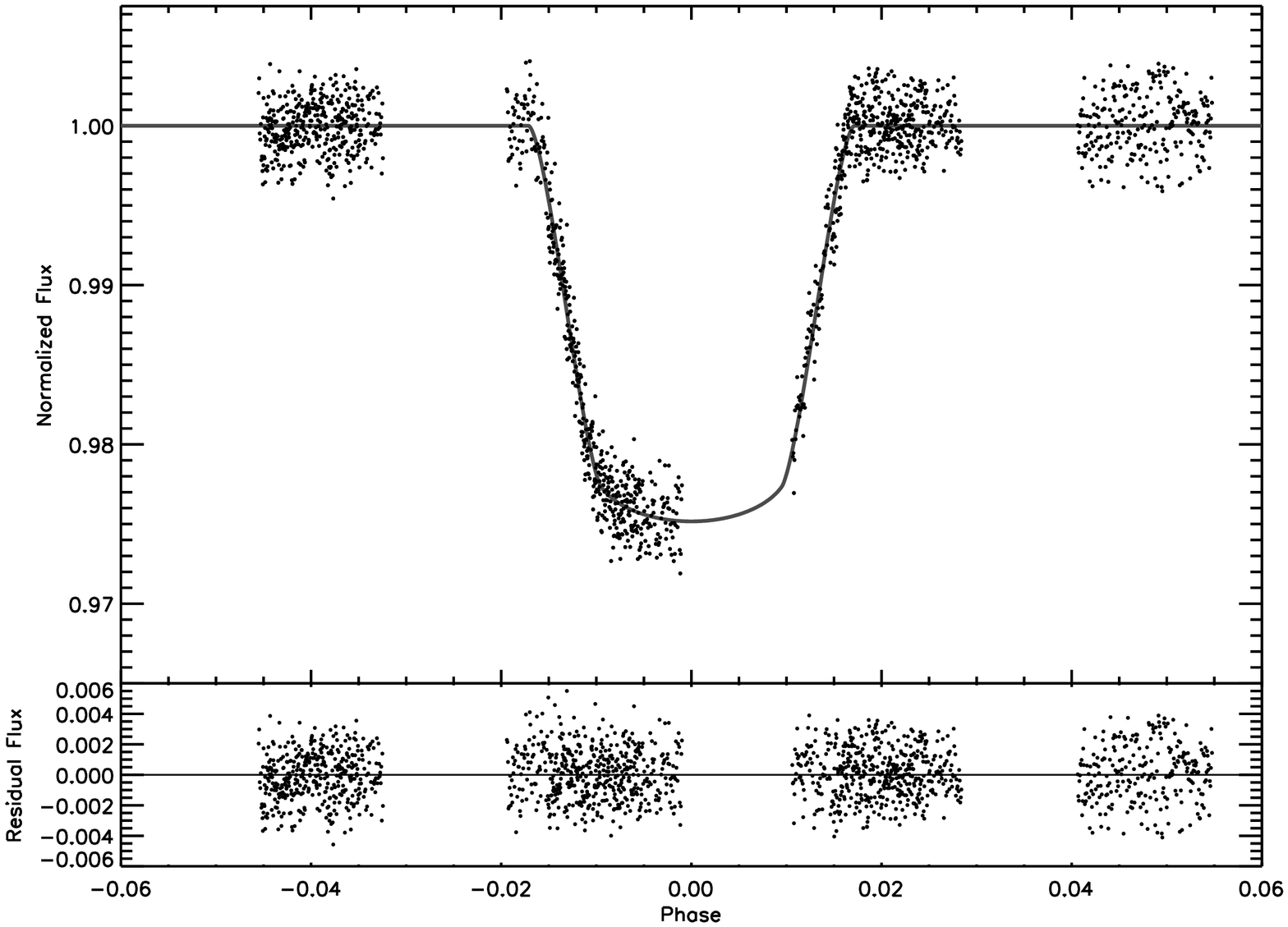}
\caption[]{\footnotesize{NICMOS filter transit photometry of both visits using the 1.6607$\mu m$ narrowband filter.  The best-fit transit light curve model is also fit (grey) along with the observed - calculated residuals of the fit (bottom panel).  The light curve has a S/N of 600 per-point and an equivalent precision of 6.3$\times10^{-4}$ in an 80 second bin.}}
\end{figure}

  The fit values for the linear limb-darkening coefficients are consistent within 1$\sigma$ of the model values (see Table 2).  We also performed fits letting either $c_3$ or $c_4$ free, with $c_2$ fixed, finding similar results indicating that the Atlas models are performing well.  As a check of our choice adopting the three parameter law, we also fit the four-parameter limb-darkening law fixing $c_2$, $c_3$, and $c_4$ to their model values and letting the $c_1$ coefficient (proportional to $\mu^{1/2}$) free.  At both 1.66 and 1.87$\mu$m, we found $c_1$ to be consistent with zero.  In addition, we tried fits with the quadratic limb-darkening law fitting both coefficients in a minimally correlated fashion \citep{2008MNRAS.390..281P}, finding equivalent transit parameters, though the three-parameter law still produced the best $\chi^2$ fits.

\begin{table*} 
\caption{System parameters of HD189733b}
\label{table:2}
\begin{centering}
\renewcommand{\footnoterule}{}  
\begin{tabular}{lllll}

\hline\hline  
Parameter  & \multicolumn{2}{c}{\underline{Individual NICMOS Fits}}     & \multicolumn{2}{c}{\underline{Joint HST \& Spitzer$^\dagger$}} \\
                         &  F166N filter           &  F187N filter         &  F166N filter  &  F187N filter  \\ 
\hline 
 central wavelength, $\lambda$ ($\mu$m)     &  1.6607  &  1.8748 \\
 wavelength range, $\Delta\lambda$($\mu$m) & 0.0170  &  0.0191 \\
planet-to-star radius contrast, R$_{pl}$/R$_{\star}$        &   0.15464$\pm$0.00051  &     0.15496$\pm$0.00028       &  0.15498$\pm$0.00035  &  0.15517$\pm$0.00019     \\ 
inclination, $i$ (deg)                 &    85.75$\pm$0.13      &     85.715$\pm$0.076    &   \multicolumn{2}{c}{85.723$\pm$0.024$^\dagger$}   \\ 
system scale, $a$/R$_{\star}$             &    8.871$\pm$0.095     &      8.854$\pm$0.062    &   \multicolumn{2}{c}{8.898$\pm$0.027$^\dagger$}  \\ 
impact parameter, $b$=$a$cos$i$/R$_{\star}$   &    0.657$\pm$0.021     &      0.661$\pm$0.013    &   \multicolumn{2}{c}{0.6636$\pm$0.0031$^\dagger$} \\ 
stellar density, $\rho_{\star}$ (g cm$^{-3}$) &   2.684$\pm$0.086       &     2.669$\pm$0.056      &  \multicolumn{2}{c}{2.709$\pm$0.025$^\dagger$} \\ 
linear limb-darkening coefficient, $c_{2}$                      &   2.195$\pm$0.058     &     1.972$\pm$0.029      &       2.1483  &  1.9508  \\ 
\hline
\end{tabular}
\end{centering}
\footnotesize{\hspace{1cm}$^\dagger$System parameter values and 1-$\sigma$ error derived from the weighted mean using: Pont et. al. (2007) HST/ACS, Desert et al. (2009a,b) Spitzer/IRAC, Knutson et al. (2009) Spitzer/IRAC, and the individual fits of this work HST/NICMOS.  The mean value was adopted for the fits.}
\end{table*}

  Given the accumulating high precision transit data on HD189733 with both HST and Spitzer, several orbital system parameters (such as $i$, $b$, and $a$/R$_{\star}$) can be determined more precisely than with just these HST/NICMOS transit observations alone.  Transit-fit values for R$_{pl}$/R$_{\star}$ are correlated to these system parameters, which can adversely affect a proper comparison of the planetary radii across different wavelengths, when searching for atmospheric signatures, if significantly deviant values are used.  Therefore, it is advantageous to fit for R$_{pl}$/R$_{\star}$ using the best system parameters available, and also necessary that the parameters are consistent across different studies when comparing the planetary radii between multiple observations.  In Figure 3, we plot the measured orbital system parameters of HD189733 from \cite{2007A&A...476.1347P}, \cite{2009ApJ...699..478D}, \cite{2009submittedD}, \cite{2009ApJ...690..822K}, \cite{2007AJ....133.1828W}, and this work using the individual NICMOS fits.  The results for $i$, $a$/R$_{\star}$, $b$, and $\rho_{\star}$ nearly all agree at the $\sim1\sigma$ level.  A notable exception is the 3.6$\mu m$ and 5.8$\mu m$ Spitzer results from \cite{2009ApJ...699..478D}, which show significantly deviant values in both $b$ and $a$/R$_{\star}$.  These deviant values may have been the result of occulting stellar spots during ingress or egress and were disregarded in this study (see Desert et al. 2009b for further details).  From the HST, Spitzer, and ground-based studies, we derived mean orbital system parameter values calculating a weighted mean using the different measurements.  The determined mean values and their associated error are listed in Table 2.  These values have an improved determination of $i$, $b$, $\rho_{\star}$ and $a$/R$_{\star}$ by up to a factor of seven compared to our individual NICMOS fits and were adopted for our final transit light curve fits (labeled {\it Joint HST \& Spitzer} in Table 2 and plotted in Fig. 4 and 5).  We also choose to adopt the model limb-darkening values from Table 1 for the three coefficients, minimizing the number of free fit parameters.

\subsection{Red noise estimation}
The RMS of the residuals between the best-fitting Joint HST \& Spitzer
  models and the data are 9.98$\times10^{-4}$ for the F187N filter and
  1.67$\times10^{-3}$ for the F166N filter.  These values are
  approximately 1.5 times the calculated pipeline error, which is
  dominated by photon noise.  This is a modest improvement over other
  similar NICMOS transit studies
  \citep{2009ApJ...696..241C,2009MNRAS.393L...6P}, which have generally
  found 2-3 times the photon noise limit.  We checked for the presence
  of systematic errors correlated in time (``red noise'',
  \citealt{2006MNRAS.373..231P}) using two methods.  We first checked
  that the binned residuals followed a $N^{-1/2}$ relation when
  binning in time by N points.  The presence of red noise causes the
  variance to follow a $\sigma^2=\sigma_w^2/N+\sigma_r^2$ relation,
  where $\sigma_w$ is the uncorrelated white noise component while $\sigma_r$ characterizes the red noise.  We found no significant evidence for red noise in either the F187N data or F166N data, when binning on timescales up to the ingress and egress duration.  We also used the ``prayer-bead'' method to check for red noise, which consists of a residual permutation algorithm \citep{2004A&A...424L..31M}.  In this method, the residuals of the best-fit model are shifted with wraparound over the light curve and a new fit is performed.  The structure of any time correlated noise is preserved and its effect on the uncertainty of the transit light curve parameters are revealed in the distribution of fit parameters.  The prayer-bead method also showed no significant signs of red noise, as the distribution of each of the fit parameters at both wavelengths were compatible with the uncertainties found from standard $\chi^2$ statistics.
With no detectable red noise, our S/N levels are comparable to other high S/N transit light curves, with precisions of 4.0$\times10^{-4}$ per 80 second bin for F187N and 6.3$\times10^{-4}$ per 80 second bin for F166N.

\begin{figure*}
\begin{centering}
 {\centering
  \includegraphics[width=0.615\textwidth,angle=90]{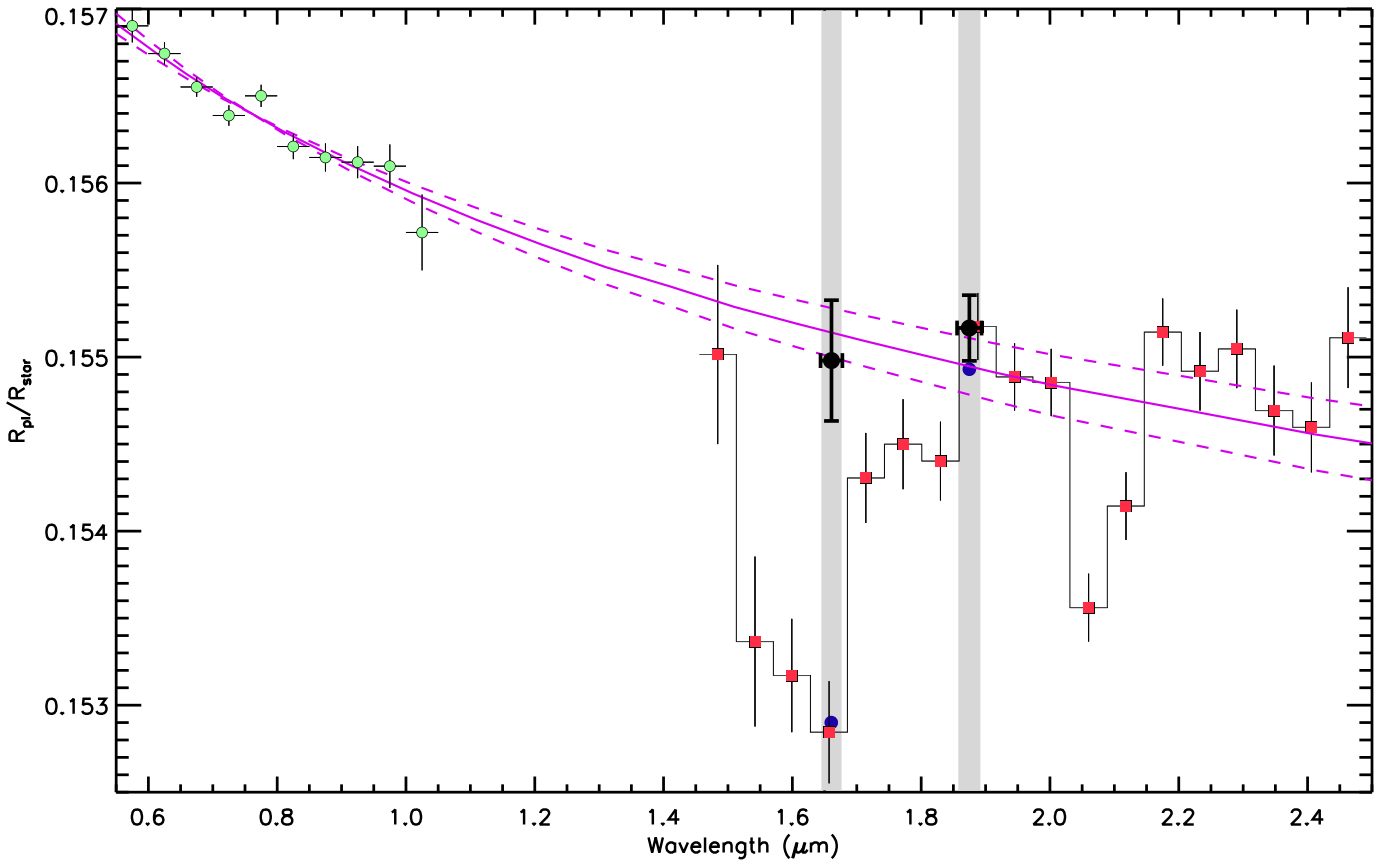}}
\caption[]{\footnotesize{Measured planetary radii for HD189733b at optical and near-infrared wavelengths.  Plotted are the results from our NICMOS narrowband photometry (black dots), along with the NICMOS grism spectrum (red squares) from Swain et al. (2008), and ACS grism spectrum (green dots) from Pont et al. (2008).  The 1-$\sigma$ error bars on the fit radii are indicated (y-axis error bars), along with the wavelength range of each observation (x-axis error bars, grey vertical bars).  Also plotted (purple) is the prediction by Rayleigh scattering due to haze from Lecavelier et al. (2008a), projected here into the near-infrared along with the 1-$\sigma$ error on the predicted slope (purple, dashed lines).  The NICMOS spectrum from Swain et al. (2008) is quoted as being uncertain in its absolute flux level by $\pm$2$\times10^{-4}$ ($\pm$0.00064 R$_{pl}$/R$_{\star}$), an offset of -0.00042 R$_{pl}$/R$_{\star}$ was applied here for comparison reasons such that the values at 1.87 $\mu m$ match.  Our 1.66$\mu m$ results are in disagreement with the both the Swain et al. spectra and the expected H$_2$O atmospheric signature (blue dots), but are in excellent agreement with the predicted planetary radii values from atmospheric haze.}}
\end{centering}
\end{figure*}

\section{Discussion}
\subsection{Expected H$_2$O atmospheric signature}
We have presented high signal-to-noise near-infrared transit narrowband photometry of HD189733 at 1.66
and 1.87 $\mu m$, with the intent of searching for a transit signature of
atmospheric water.  If atmospheric H$_2$O were the dominant absorbing
species at both of these wavelengths, we could expect significant radii
differences as H$_2$O has a strong absorption bandhead over the probed
wavelength range.  The expected difference in radii can be estimated using the difference between opacities at 1.66$\mu m$ and 1.87$\mu m$.  We calculate the effective absorption cross sections per molecule to be $\sigma_{1.66}$=8.54$\times10^{-24}$ and $\sigma_{1.87}$=1.79$\times10^{-21}$ (cm$^2$ molecule$^{-1}$), respectively, using the high temperature line list of \cite{2006MNRAS.368.1087B}, a Voigt broadening line profile, a temperature of 1200 K, and the response through our two narrowband filters.  The variation in apparent planetary
radius then follows from Eq. 2 of \citep{2008A&A...481L..83L} giving, 
\begin{equation} \Delta R_p=H\ln\frac{\sigma_{1.87}}{\sigma_{1.66}},  \end{equation}
where $H$ is the atmospheric scale height.  
$H$ is given by
H = $kT/\mu g$, where $\mu$ is the mean mass of the atmospheric particles
taken to be 2.3 times the proton mass, $T$ the temperature, and $g$ the surface gravity.
Assuming $R_{\star}=0.766$ $R_{\odot}$, $R_{pl}=1.155$ $R_{Jup}$, $M_{pl}=1.144$ $M_{Jup}$, and a temperature of 1200 K, we find $H/R_{\star}$=0.00038 which gives an expected radius difference of 0.0020 R$_{pl}$/R$_{\star}$.  Our HST NICMOS observations show only a radius contrast of 0.00019$\pm$0.00040 R$_{pl}$/R$_{\star}$ between the two wavelengths, a factor of 10 lower and 4.6$\sigma$ away from the expected value.
If H$_2$O was the dominant opacity source in HD189733b's atmosphere at
these altitudes, it would have been easily detected by our observations.

\subsection{Comparisons with previous results}
\cite{2008Natur.452..329S} used HST NICMOS to produce a near-IR low resolution grism transmission spectrum of HD189733b, which appears to show two strong atmospheric features at H$_2$O and CH$_4$.  This grism spectrum encompasses our wavelength range (see Fig. 6) and appears to show a strong H$_2$O absorption molecular bandhead feature between 1.5 and 2.0 $\mu m$.  Between 1.66 and 1.87 $\mu m$, \cite{2008Natur.452..329S} found a radii contrast of 0.00233$\pm$0.00035 R$_{pl}$/R$_{\star}$.  This radii contrast is significantly larger than the one we observe, with our results failing to reproduce the Swain et al. value at a 5$\sigma$ confidence level.

Given the large disagreement between our results and \cite{2008Natur.452..329S}, we find two scenarios that could possibly explain the conflicting results.  Either the planet's transmission spectrum is variable, or residual systematic errors still plague the edges of the Swain et al. spectrum.  Both \cite{2008Natur.452..329S} and our results correct for stellar starspot activity, ruling out stellar variability as the underlying cause.  In addition, the discrepancy between the two results is too large to attribute to random statistical fluctuations.  Variability in the planet's transmission spectrum by dynamic weather processes would seem a plausible scenario, especially given the seemingly variable emission spectrum results from Spitzer \citep{2007ApJ...658L.115G,2008Natur.456..767G}.  However, we find the transmission variability scenario difficult to substantiate, particularly with the large systematic errors inherent with NICMOS grism observations, which can compromise precision photometric work.  
In order for the variability scenario to be viable, the optically thick absorbing haze would have to clear out of the atmosphere, over the entire limb, down to altitudes at least 6.7 scale heights lower ($\sim$1200 km) to reveal the full extent of the H$_2$O feature.  The haze would also have to clear for only the single epoch of the Swain et al. (2008) HST visit, but not during the seven epochs of the HST visits between Pont et al. (2008) and this work.  Likely, the variability scenario is also dependent upon the haze composition, and the transmission spectrum would be difficult to vary with non-condensate Rayleigh scattering candidates such as H$_2$ \citep{2008A&A...481L..83L}.  We note that in our HST visits spanning five months, no significant visit-to-visit planet radius differences were observed at either wavelength, after correction for non-occulted stellar spots.  At 1.66 $\mu m$ we find radii of 0.1545$\pm$0.0004 and 0.1551$\pm$0.0006 R$_{pl}$/R$_{\star}$ for visits \#2 and 4, respectively, while at 1.88 $\mu m$ we find radii of 0.1549$\pm$0.0003 and 0.1558$\pm$0.0004 R$_{pl}$/R$_{\star}$ for visits \#1 and 5, respectively.

 HST NICMOS slit-less grisms have a number of instrument related systematic effects that appear in precision photometry.  These effects include detector wavelength dependancies such as the flat field, sensitivity, PSF, and resolution which interact in a complex manner with telescope's PSF stability and pointing accuracy.  With these effects, typical transit programs with HST NICMOS grisms tend to exhibit large intra-orbit flux variations as the telescope thermally relaxes from day-to-night and orbit-to-orbit flux variations largely dependent on the grating filter wheel position and spectral trace (see Fig. 1 \& 2 of \citealt{2009ApJ...696..241C}).  In addition, the low resolution Nic 3 camera used for grism observations is highly under-sampled leading to large intra-pixel sensitivities, though these can be largely averaged out by defocusing the detector.  The importance of systematic errors resulting from the grating filter wheel position on HD189733b's spectrum can be seen in the emission spectrum from \cite{2009ApJ...690L.114S}.  This spectrum shows nonphysical results (with negative planet fluxes) at the blue-edge of the spectrum.  At the edges, the spectral trace has the largest orbit-to-orbit deviations, causing greater systematic errors.  The HD189733b NICMOS transmission and emission spectra were both observed and reduced in much the same manner by Swain et al., thus the edges of the transmission spectra should also be suspect of underestimated uncertainties and greater residual systematic errors.

In a separate NICMOS spectroscopic study of \object{GJ436b} by Pont et al. (2008), variations at the $4\times10^{-4}$ flux level in the transmission spectrum were also found, similar in amplitude to the Swain et al. spectral features of HD189733b.  However, given the level of measured systematic errors, Pont et al. (2008) chose not to interpret the GJ436b variations as due to real planetary transmission features, but rather as systematic errors.  These errors exhibited a wavelength dependence that was time dependent, with spectral features slowly shifting in wavelength with time, as expected from the HST thermal focus variation systematic effect.  While Pont et al. (2008) and Swain et al. (2008) used different NICMOS grisms, both corrected for systematic errors in a similar manner adopting a multilinear decorrelation against external variables including the position, rotation, and width of the spectral trace.  Thus, similar systematic errors between the two studies may be expected.  A systematic error of $4\times10^{-4}$ in flux would correspond to an error of $\sim$0.0013 R$_{pl}$/R$_{\star}$ at a given wavelength in HD189733b and $\sim$0.0018 R$_{pl}$/R$_{\star}$ when comparing the difference between two wavelengths, possibly accounting for the features in the Swain et al. (2008) spectrum.   

With NICMOS grisms, wavelength shifts and a variable PSF with time can change the flux per pixel with time, which becomes compounded by pixel-to-pixel and sub-pixel response variations which are not accurately known.
Observing with narrowband photometric filters mitigates these wavelength dependent and PSF effects, eliminating the need to correct for the systematic errors shown by the grism spectra (a main motivation behind our original selection of narrowband filters).  Given our robust (decorrelation free) photometry and the suspect edges of the transmission spectra, the differences between our results and \cite{2008Natur.452..329S} would clearly seem to favor grism-related systematic errors as the principal cause.
\begin{figure}
\begin{centering}
 {\centering
  \includegraphics[width=0.35\textwidth,angle=90]{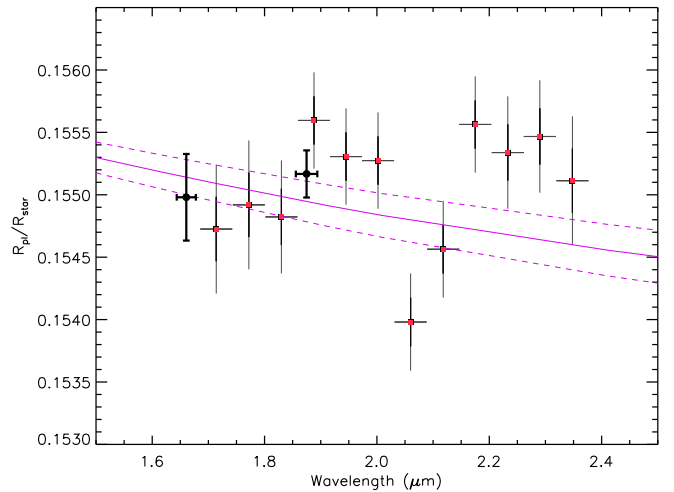}}
\caption[]{\footnotesize{Similar to Fig. 6, except the Swain et al. (2008) data are plotted at the original unshifted level and the points at the ends of the spectrum (likely affected by systematic errors) are removed.  Both 1$\sigma$ error-bars (thick) and 2$\sigma$ error-bars (thin) on the Swain et al. spectra are also shown to illustrate that it is consistent at that level with Rayleigh scattering by condensate haze, below 2 $\mu m$.}}
\end{centering}
\end{figure}

\subsection{Atmospheric haze in the near-infrared}
Although our results on the planet-to-star radius ratios measured at 1.66~$\mu$m and 1.87~$\mu$m are significantly different from the \cite{2008Natur.452..329S} measurements at the same wavelengths, our results are still consistent at both wavelengths with the extrapolation of measurements at optical wavelengths assuming Rayleigh scattering.  In effect, the accurate measurements of \cite{2008MNRAS.385..109P} showed a significant decrease of the planet radius toward longer wavelengths that was interpreted as the signature of absorption by haze.  \cite{2008A&A...481L..83L} showed that the radius variation from 0.55 to 1.05~$\mu$m is a signature of the Rayleigh scattering with a decrease of the cross-section following a $\lambda^{-4}$ power law. This law can be extrapolated to 1.87~$\mu$m. We find that both our measurements at 1.66~$\mu$m and 1.87~$\mu$m 
are in excellent agreement with the Rayleigh scattering prediction (Fig.~6).  As shown in \cite{2008A&A...481L..83L}, the slope of the radius as a function of the wavelength in the Rayleigh regime is characteristic of the atmospheric temperature.  The slope for extreme temperatures of 1190 and 1490\,K are plotted in Fig.~6 (dotted purple lines).  We can see that the smaller slope corresponding to the lowest temperatures is slightly favored, in particular by the 1.87~$\mu$m measurement.

We performed a global fit to the present new measurements together with the \cite{2008MNRAS.385..109P} published values. 
With the hypothesis of pure Rayleigh scattering, the data are well fitted with a reduced $\chi ^2$ of 1.1 despite the extended width of the spectral domain spread over a factor 4 in wavelength.
We find that the temperature given by the Rayleigh power law is 1280$\pm$110\,K. 
This temperature agrees with Spitzer 8 and 24~$\mu m$ phase curve measurements, which also find similar temperatures of $(T_{max}+T_{min})/2$ = 1134 and 1102\,K respectively \citep{2009ApJ...690..822K}. 
Relaxing the Rayleigh hypothesis and using Mie scattering from 0.55~$\mu$m to 1.87~$\mu$m, we obtain new constraints on the particle size and their optical properties (see \citealt{2008A&A...481L..83L}), with the fit to the data slighly improved.
Mie scattering implies a larger possible temperature range of 1170 to 1870\,K (includes 1$\sigma$ error) because of a less steep extinction variation with wavelength in the near infrared.
The proposed scenario with MgSiO$_3$ still holds, and we find that the particle maximum size must be in the range 0.009\,$\mu$m to 0.086\,$\mu$m.
Recall that because the Rayleigh scattering cross-section is proportional to the particle size, $a_{\rm part}$, to the power of 6 ($\sigma\propto a_{\rm part}^6$), the size range given above is the size of the largest particles in the size distribution.  

Finally, the interpretation of the present transit absorption signatures in terms of Rayleigh scattering at $\lambda\la 2$\,$\mu$m is consistent with the measurements of emission spectra which do not show signatures of haze absorption. This is explained by the fact that the slant transit geometry probes atmospheres at high altitude and is more sensitive to low absorption cross-sections, while the normal viewing geometry of emission spectra probes deeper in the 
atmosphere \citep{2005MNRAS.364..649F}. Though more uncertain in the absolute planet-to-star radius contrast, the \cite{2008Natur.452..329S} transmission spectrum can also be considered as consistent with the haze interpretation if the extreme parts of their absorption spectra are disregarded (Fig.~7).

\section{Conclusions}
Narrow-band NICMOS photometry can provide a robust method for obtaining precision near-infrared photometry.  With transiting planets, this can lead to secure near-infrared planetary radii where important molecular and condensate atmospheric planetary signatures can be found.  With two carefully selected narrow-band filters, we are able to provide stringent constraints on the presence of H$_2$O absorption features.  We find that upper-atmospheric haze provides sufficient opacity in slant transit geometry to obscure the near-IR H$_2$O molecular signatures below 2$\mu m$.  While absorption due to H$_2$O was reported by \cite{2008Natur.452..329S}, we rule out any such feature at 5$\sigma$ and find that their observed feature is likely an artifact of residual systematic errors compromising the spectral edges. 
The planetary radii values from \cite{2008MNRAS.385..109P} and this work indicate that Rayleigh scattering dominates the broadband transmission spectrum from at least 0.5 to 2 $\mu m$, with sub-micron MgSiO$_3$ haze particles a probable candidate \citep{2008A&A...481L..83L}.  The transition wavelength between the transmission spectra being dominated by haze particles and molecular features would seem to be between 2 and 3 $\mu m$, as the IRAC Spitzer photometry between 3 and 8 microns all show planet radii in excess of those predicted by Rayleigh scattering \citep{2009ApJ...699..478D, 2009submittedD}.
To be clear, while our observations rule out detectable quantities of H$_2$O in the near-IR at these high altitudes, the presence of methane in the Swain et al. transmission spectrum is still possible, though it needs verification, as the 2.2 to 2.4 $\mu m$ region appears to show an excess absorption over the Rayleigh prediction. 

Spectroscopic characterization of transiting planets
with JWST NIRCam and NIRSpec are anticipated
to make significant contributions to exoplanet science \citep{2009arXiv0903.4880D}.
These instruments have similar detectors and
observing modes as HST NICMOS, with NIRCam
also containing similar narrow-band filters to those used here.  
While the systematic errors observed with HST and Spitzer should be greatly reduced
in JWST, the correspondingly greater precisions that
will be attempted, as super-Earths are observed, ensure
that at some level these systematic errors will likely
remain.  Strategic use of narrowband filters can provide
robust measurements with which to compare atmospheric features
seen in transmission or emission spectra, checking both observing
methods at key wavelengths.  As no upcoming
instrument will soon be able to compete with JWST,
such independent methods may be the only way to confidently
verify groundbreaking results in super-Earth spectra requiring
significant systematic error corrections.


\begin{acknowledgements}
D.K.S. is supported by CNES.  G.E.B. was supported for this work by NASA through grant GO-11117 to
the Univ. of Arizona from the STScI.
G.W.H. acknowledges long-term support from NASA, NSF, Tennessee State University,
and the State of Tennessee through its Centers of Excellence Program.
This work is based on observations
with the NASA/ESA Hubble Space Telescope, obtained at the
Space Telescope Science Institute (STScI) operated by AURA, Inc.
We thank the referee F. Pont for the valuable insights into NICMOS grism errors.  
D.K.S. and G.E.B. would like to thank M. Rieke for valuable discussions 
regarding the NICMOS instrument.  D.K.S. would also like to thank the helpful and knowledgeable staff
at STScI for their input in the planning and execution of the observations.
\end{acknowledgements}

\bibliographystyle{aa} 
\bibliography{Sing} 

\end{document}